\documentstyle[aps,preprint,pra]{revtex}

\renewcommand{\ref}[1]{\raisebox{.6ex}{[#1]}}

\newcommand{\be}{\begin{equation}}
\newcommand{\ee}{\end{equation}}

\newcommand{\bea}{\begin{eqnarray}}
\newcommand{\eea}{\end{eqnarray}}

\newcommand{\ba}{\begin{array}}
\newcommand{\ea}{\end{array}}

\begin{document}



\title{ Two Stages in Evolution of Binary Alkali BEC Mixtures 
          towards Phase Segregation  }

\author{           P. Ao \\
         Department of Theoretical Physics  \\
         Ume\aa{\ }University, 901 87 Ume\aa,  Sweden  \\ 
                  and \\
                 S.T. Chui    \\
          Bartol Research Institute \\
         University of Delaware, Newark, DE 19716, USA \\
          }

\maketitle


\begin{abstract}
The quantum analogy of the usual spinodal decomposition
is explored for the recently achieved binary alkali Bose-Einstein condensates
mixture.
We conclude that an analogy is possible within the formulation of coupled 
non-linear Schr\"odinger equations, 
and find that the quantum spinodal decomposition consists of two stages.
The non-equilibrium stage I is dominated by the fastest growth mode,
associated with a characteristic length.
Expressions for both time and length scales in the stage I are obtained.
The state II is a relaxation process of approaching equilibrium, dominated
by the slowest mode.
We propose that for this state the slow evolution 
towards the phase segregation 
is due to the Josephson effect between different domains of same condensate, 
and its time scale is estimated.

\noindent
PACS${\#}$:  03.75.Fi; 64.75.+g 
 
\end{abstract}

 

\section{Introduction}

Spinodal decomposition in a binary-solution system is a
typical example of phase ordering dynamics:
the growth of order through domain coarsening 
when a system is quenched from the homogeneous phase into 
a broken-symmetry phase \cite{cahn}.
Systems quenched from a disordered phase into
an ordered phase do not order instantaneously.
Instead, different length scales set in as the domains form and 
grows with time, 
and different broken symmetry phases compete to select
the equilibrium state.
In the dynamical equation, the Cahn-Hilliard equation,
for the spinodal decomposition, the diffusion constant of the material
plays a decisive role and determines both the length and time scales.
It appears that a system without the diffusion constant may not show
all the main features of the spinodal decomposition.
With the recent
realization of the binary Alkali Bose-Einstein condensates (BEC's) \cite{jila1}
we demonstrate in the present article that indeed it is possible
to have a spinodal decomposition without the diffusion constant and that
this may have  been realized experimentally.
The binary alkali BEC mixtures provide  new systems to non-equilibrium
phenomena in new parameter regimes. 
Its mathematical description is
simple enough that a theoretical description of
the whole process is feasible. 
A direct comparison between theoretical calculations and
experimental observations can be made.
Here, we shall study the gross 
features in the dynamical evolution process, 
starting from the homogeneous unstable state.
To differentiate the present situation from the usual ones, we shall call the 
present one the quantum spinodal decomposition, and the previous ones
the classical spinodal decomposition.

In a classical spinodal decomposition process, 
the particle number for each specie is conserved separately.
The process can be classified into two stages, the initial stage 
of fast growth from the homogeneous unstable state
and the late stage towards to the equilibrium of true ordering.
The initial stage is dominated by the fastest growth mode determined by 
the dynamical equation. It is a highly non-equilibrium process. 
There is a length scale associated with this time scale, which 
gives the characteristic domain size after this initial stage.
In the late stage the domains grow and merge, 
and their numbers becomes smaller and smaller. 
This is a relaxation process towards the equilibrium, dominated by the 
slowest time scale. For an infinitely large system, 
the time scale may be infinity. Hence the system may never 
achieve true equilibrium.
We shall show below that the quantum spinodal decomposition
occurring in the binary BEC mixtures shares all the features of the 
classical spinodal decomposition. 
The main difference is that the size of  
binary BEC mixtures is finite. Therefore
it is possible to achieve true equilibrium within a finite time scale. 
To further differentiate the present process from the classical one,
we shall call the initial stage in the 
dynamical evolution of the binary BEC mixtures
the stage I, and call the second stage 
towards equilibrium the stage II.

In the following, we first formulate the problem in Sec. II.
The coupled nonlinear Schr\"odinger equations and the parameter regime
to realize the quantum spinodal decomposition will be specified.
In Sec. III the stage I is studied in detail.
The fastest growth mode is explicitly given.
Interspersed regions or domains of coexisting condensate 1 and 2
are formed at this stage, characterized by a length scale. 
Stage II is studied in Sec. IV. 
We first demonstrate that the scenarios in classical spinodal 
decompositions are not possible here.
We propose and analyze a new mechanism for the approach towards
equilibrium: the Josephson effect between different domains of same
condensate. 
This is in accordance with the present quantum spinodal decomposition 
concept.
A comparison to experimental situations is discussed in 
Sec. V. We conclude there that the quantum spinodal decomposition
can be realized.
We summarize in Sec. VI.

\section{ Formulation of Problem }

We start from the Hamiltonian formulation of a binary BEC mixture at zero 
temperature:
\bea
   H & = & {\ } \int d^3 x \left[ \psi^\ast_1(x) 
       \left( \frac{ - \hbar^2\nabla^2 }{2m_1} \right) \psi_1(x) 
       + \psi^\ast_1(x) U_1(x) \psi_1(x) \right]  \nonumber \\
     & &  +  \int d^3 x \left[ \psi^\ast_2(x) 
       \left( \frac{ - \hbar^2\nabla^2 }{2m_2} \right) \psi_2(x) 
       + \psi^\ast_2(x) U_2(x) \psi_2(x) \right]  \nonumber \\
     & & + \frac{G_{11}}{2} \int d^3 x \; \psi^\ast_1(x) \psi_1(x) \,
                                 \psi^\ast_1(x) \psi_1(x) \nonumber \\
     & & + \frac{G_{22}}{2} \int d^3 x \; \psi^\ast_2(x) \psi_2(x) \,
                                 \psi^\ast_2(x) \psi_2(x) \nonumber \\
    & &  + G_{12} \int d^3 x \; \psi^\ast_1(x) \psi_1(x) \,
                                 \psi^\ast_2(x) \psi_2(x)  \; .
\eea  
Here $\psi_j$,  $m_j$,  $U_j$ with $j=1,2$ are
the effective wave function,  the mass, 
and the trapping potential of the $j$th condensate.
The interaction between the $j$th condensate atoms is specified by $G_{jj}$,
and that between 1 and 2  by $G_{12}$.
In the present paper all $G's$ will be taken to be positive.
The corresponding time dependent equations of motion 
are the well-known  non-linear Schr\"odinger equations \cite{nlse},
obtained here by minimization of the action, 
$ S = \int dt \{ \sum_{j=1,2}
       \psi_j^{\ast} i \hbar \frac{\partial }{\partial t} \psi_j - H \} $, 
\bea
    i \hbar \frac{\partial }{\partial t}  \psi_1(x,t) & = &  
       - \frac{\hbar^2}{2m_1} \nabla^2 \psi_1(x,t) 
       + (U_1(x) - \mu_1 ) \psi_1(x,t)
       + G_{11} |\psi_1(x,t)|^2 \psi_1(x,t)  \nonumber  \\
    & & + G_{12} |\psi_2(x,t)|^2 \psi_1(x,t) \; ,
\eea
and 
\bea
    i \hbar \frac{\partial }{\partial t} \psi_2(x,t) & = & 
      - \frac{\hbar^2}{2m_2} \nabla^2 \psi_2(x,t) 
      + (U_2(x) - \mu_2 ) \psi_2(x,t)
      + G_{22} |\psi_2(x,t)|^2 \psi_2(x,t)  \nonumber \\ 
    & & + G_{12} |\psi_1(x,t)|^2 \psi_2(x,t)  \; .
\eea
  The Lagrangian multipliers, the chemical potentials $\mu_1$ and $\mu_2$, 
are fixed by the relations
$
   \int d^3 x  |\psi_j(x,t)|^2  =  N_j \; , j =1,2
$,
with $N_j$ the number of the $j$th condensate atoms.
Eq. (2) and (3) are mean field equations, since we treat the effective 
wave functions as $c$ numbers, corresponding to
the Hartree-Fock-Bogoliubov approximation.

Experimentally, the trapping potentials $\{ U_j \}$ are
simple harmonic in nature.
For the sake of simplicity and to illustrate the physics
we shall consider a square well trapping potential $U_j = U$:
zero inside and large (infinite) outside,
unless otherwise explicitly specified.
We will come back to this question in the discussion 
of experimental feasibility.

We shall consider the strong mutual repulsive regime 
\be
   G_{12} > \sqrt{ G_{11} G_{22} }   \; .
\ee
In this regime the equilibrium state for two Bose-Einstein condensates
is a spatial segregation of two condensates, 
where two phases, the weakly and strong 
segregated phases, characterized by the healing length and the penetration 
depth, have been predicted \cite{ac}.
We shall  use Eq. (2) and (3) under the condition (4) to study a highly 
non-linear dynamical process: 
The two condensates are initially in a homogeneously mixed state,
then eventually approach to the phase segregated state.
In the same mean field manner as in Ref.\onlinecite{ac}, 
we find that this dynamic
process can be classified into two main stages:
The initially high non-equilibrium dynamical growth in stage I, where the
dynamics is governed by the fastest growth mode, and the stage II of 
approaching to equilibrium where the dynamics 
is governed by the slowest mode.
The stage II is the typical of a relaxation process near equilibrium.
However, we shall show again it is governed by a quantum effect, namely,
the Josephson effect.

\section{ Stage I:  Fastest Growth Mode  }

With the square well trapping potential specified in Sec. II,
the coupled non-linear Schr\"odinger equations
have an obvious homogeneous solution:
Inside the trap the condensate densities $|\psi_j|^2 = \rho_{j0} $,
$
   \rho_{j0} =  {N_j }/{V} \; , 
$
with $V$ the volume of the square well potential trap,
and the chemical potentials
$
   \mu_1 = G_{11} \rho_{10} + G_{12}\rho_{20}   
$
and
$ 
   \mu_2 = G_{22} \rho_{20} + G_{12}\rho_{10}   
$.
This is the initial condition of the present problem.
It is known that for a large enough mutual repulsive interaction, 
that is, if
$
   G_{12} > \sqrt{ G_{11} G_{22} }  \; ,
$
this initial state is not the ground state. 
Rich physics has been displayed by various theoretical studies \cite{pu,ca}.
Among many unstable and growing modes in this parameter regime,
we will find the fastest growth mode in the initial stage of the process 
towards to the equilibrium state.

To look for the fastest growth mode out of the homogeneous state,
we start with small fluctuations from the homogeneous state.
This is consistent with the usual stability analysis \cite{pu}.
Our approach here is to emphasize the connection with the
physics of the classical spinodal decomposition and the role played by the 
Josephson relationships. Define
\be
   \psi_1(x,t) = \sqrt{\rho_1(x,t) } \; e^{i\theta_1(x,t) } \; ,
\ee 
and 
\be
   \psi_2(x,t) = \sqrt{\rho_2(x,t) } \; e^{i\theta_2(x,t) } \; .
\ee
and define the density fluctuations $\delta \rho_j = \rho_j - \rho_{j0} $
and the phase fluctuations  $ \theta_j $, and 
assume they are small: 
$  |\delta\rho_j | / \rho_j \; , \; |\theta_j | << 1 $.
The definition of the phase fluctuations here has made use of
the implicit assumption that there is no average current.
Then, to the linear order, we have from Eq. (2,3),
for condensate 1,
\bea
    & & \delta \dot{\rho}_{1} + 
     \rho_{10} \frac{ \hbar }{m_1 } \nabla^2 \theta_1  =   0  \; , \\
    & & \hbar \dot{\theta}_1  =  \frac{ \hbar^2 }{4m_1 } 
        \frac{ \nabla^2 \delta\rho_1 }{\rho_{10} } - G_{11} \delta\rho_1 
          - G_{12} \delta\rho_{2} \; .
\eea
In terms of hydrodynamics, 
Eq.(7) is the continuity equation, and Eq.(8) is the Bernoulli equation, 
with the first term at the right hand side 
as the so-called `quantum pressure'.
Similarly, for condensate 2 we have 
\bea
   & & \delta \dot{\rho}_{2} + 
       \rho_{20} \frac{ \hbar }{m_2 } \nabla^2 \theta_2  =  0 \; , \\
   & & \hbar \dot{\theta}_2  =  \frac{ \hbar^2 }{4m_2 } 
        \frac{ \nabla^2 \delta\rho_2 }{\rho_{20} } - G_{22} \delta\rho_2 
          - G_{12} \delta\rho_{1} \; .
\eea
Eliminating the phase variables from Eqs. (7-10), we have
\be
   \frac{ \partial ^2 }{\partial t^2 } 
     \left( \begin{array}{c} 
            \delta\rho_1 \\
            \delta\rho_2  \end{array}   \right)
   = \left( \begin{array}{cc} 
     -\frac{\hbar^2}{4m_1^2}\nabla^4 + \frac{\rho_{10}}{m_1} G_{11}\nabla^2 &
                  \frac{ \rho_{10}}{m_1} G_{12 } \nabla^2  \\
                  \frac{ \rho_{20}}{m_2} G_{12 } \nabla^2 &
     -\frac{\hbar^2}{4m_2^2}\nabla^4 + \frac{\rho_{20}}{m_2} G_{22}\nabla^2
            \end{array}  \right) 
      \left( \begin{array}{c} 
            \delta\rho_1 \\
            \delta\rho_2  \end{array}   \right)  \; .
\ee
We look for the solution of the form
\[
   \left( \begin{array}{c} 
            \delta\rho_1 \\
            \delta\rho_2  \end{array}   \right)
   = \left( \begin{array}{c} 
            A \\
            B  \end{array} \right) e^{i ({\bf q}\cdot {\bf r} - \omega t)} \; ,
\]
with $A,B$ constants.
Eq. (11) then becomes 
\[
  \left( \begin{array}{cc} 
     \frac{\hbar^2}{4m_1^2} q^4 + \frac{\rho_{10}}{m_1} G_{11} q^2 
       -  \omega^2 &
                  \frac{ \rho_{10}}{m_1} G_{12 } q^2  \\
                  \frac{ \rho_{20}}{m_2} G_{12 } q^2 &
     \frac{\hbar^2}{4m_2^2} q^4 +\frac{\rho_{20}}{m_2} G_{22} q^2
        - \omega^2    \end{array}  \right) 
   \left( \begin{array}{c} 
            A \\
            B  \end{array}   \right) = 0  \;  .
\]
In order to have a non-zero solution for $A$ and $B$, the determinant 
must be zero:
\be
  \det 
   \left( \begin{array}{cc} 
     \frac{\hbar^2}{4m_1^2} q^4 + \frac{\rho_{10}}{m_1} G_{11} q^2 
       -  \omega^2 &
                  \frac{ \rho_{10}}{m_1} G_{12 } q^2  \\
                  \frac{ \rho_{20}}{m_2} G_{12 } q^2 &
     \frac{\hbar^2}{4m_2^2} q^4 + \frac{\rho_{20}}{m_2} G_{22} q^2
        - \omega^2    \end{array}  \right)  = 0  \; ,
\ee
which determines the dispersion relation between the frequency and the wave
number.
One can easily check that for the one component condensate case, by putting
$G_{12} = 0$, 
Eq.(12) indeed gives the usual phonon spectrum.
In the present binary BEC mixture case, we have
\be
   \omega_{\pm}^2 
    = \frac{q^2}{2} \left[  b_{11} + b_{22}  \pm 
        \sqrt{(b_{11} - b_{22} )^2 + 4b_{12}b_{21} } \right] \; ,
\ee
with
\bea
   b_{11}  & = &  \frac{\hbar^2}{4m_1^2 } q^2 
              + \frac{\rho_{10}}{m_1} G_{11} \; , \; 
   b_{12}    =  \frac{\rho_{10}}{m_1} G_{12}  \; ,\nonumber \\
   b_{21 } & = & \frac{\rho_{20}}{m_2} G_{12} \; , \;  
   b_{22}    =  \frac{\hbar^2}{4m_2^2 } q^2 
              + \frac{\rho_{20}}{m_2} G_{22} \; . \nonumber
\eea
Obviously, both $\omega_{+}^2$ and $\omega_{-}^2$ are real as given by Eq.(13).
They give the two phonon velocities in the binary BEC mixture:
\[
  c_{\pm} = \sqrt{ \left[ \rho_{10} G_{11}/ m_1 + \rho_{20} G_{22} / m_2
 \pm \sqrt{ (\rho_{10} G_{11}/ m_1  - \rho_{20} G_{22} / m_2 )^2 
            + 4 \rho_{10} \rho_{20} G_{12}^2/(m_1 m_2) } \right] /2 } \; .
\] 
However, if 
\be
   b_{11} b_{22} - b_{12}b_{21} < 0 \; ,
\ee
the branch $\omega_{-}^2$ can be negative. This implies an imaginary frequency
$\omega_{-}$ and an imaginary phonon velocity,
which shows that the initial homogeneously mixed state is unstable.
One can verify that the sufficient condition for the inequality  Eq.(14)
to hold for small enough wave number $q$ 
is the validity of the inequality Eq.(4).
The modes defined by Eq. (13) with imaginary frequencies will then 
grow exponentially with time.
Unlike the usual situation near equilibrium, 
this growth from the present 
non-equilibrium homogeneously mixed state 
will be dominated by the fastest growth mode.
This is precisely the same case as in the initial stage of 
the classical spinodal decomposition.

Now we look for the fastest growth mode.
It is equivalent to find the most negative value of $\omega_{-}^2$.
The condition $\partial \omega_-^2 /\partial ( q^2 ) = 0$ leads to
\bea
   & & \frac{ \left[ \left( b_{11} - b_{22} \right)^2 
        + \left(  b_{11} - b_{22}  \right)
       \left(\frac{\hbar^2}{4m_1^2 }-\frac{\hbar^2}{4m_2^2 } \right)q^2_{max}
       + 4 \frac{\rho_{10}}{m_1}\frac{\rho_{20}}{m_2} G_{12}^2 \right]^2 }
    { \left[ \left(  b_{11} - b_{22}  \right)^2 
         + 4 \frac{\rho_{10}}{m_1}\frac{\rho_{20}}{m_2} G_{12}^2 \right] }
     \nonumber \\
   & & =  \left[ 2 q^2_{max} 
     \left( \frac{\hbar^2}{4m_1^2 } + \frac{\hbar^2}{4m_2^2 } \right)
    + \frac{\rho_{10}}{m_1} G_{11} + \frac{\rho_{20}}{m_2}G_{22} \right]^2 \; ,
\eea
with 
\[
  b_{11} - b_{22} = 
   \left( \frac{\hbar^2}{4m_1^2 }
                              - \frac{\hbar^2}{4m_2^2 } \right)q^2_{max} 
                         + \frac{\rho_{10}}{m_1} G_{11}  
                         - \frac{\rho_{20}}{m_2} G_{22}  \; .
\]
Eq.(15) is actually a quartic equation for $q^2_{max}$, 
whose algebraic solutions can be found by the standard way.
The corresponding imaginary frequency branch is 
\be
   \omega_{-max}^2 = - \frac{q_{max}^4 }{2} 
    \left[ \frac{\hbar^2}{4m_1^2 } + \frac{\hbar^2}{4m_2^2 }
     - \frac{ \left( b_{11} - b_{22}  \right)
            \left(\frac{\hbar^2}{4m_1^2 } - \frac{\hbar^2}{4m_2^2 }\right) }
    { \sqrt{ \left( b_{11} - b_{22}  \right)^2
         + 4 \frac{\rho_{10}}{m_1} \frac{\rho_{20}}{m_2} G_{12}^2 } }
     \right] \; .
\ee
The detailed general expressions for both $q_{max}$ and 
$\omega_{-max}$ are not physically 
transparent and  particularly illustrative. We will not explore them here.

To get a better understanding of the physical implications of 
Eqs. (15,16), we consider a special case relevant to recent experiments where
particles of the two condensates have the same mass,
$ m_1 = m_2 = m$.
In this case we find the wavenumber corresponding to the most negative
$\omega_{-max}^2$ is 
\be
   q_{max}^2 = \frac{m}{\hbar^2 } 
     \left[ \sqrt{ ( \rho_{10} G_{11} + \rho_{20} G_{22} )^2
            + 4 {\rho_{10}\rho_{20} } G_{11}G_{22} 
                \left( \frac{G_{12}^2 }{G_{11}G_{22} } - 1 \right)  }
          -  (\rho_{10} G_{11} + \rho_{20} G_{22} ) \right] \; ,
\ee
and 
\be
   \omega_{-max} = - i \frac{\hbar }{2m} q_{max}^2   \; .
\ee

The physics implied in Eqs. (15,16) or Eqs. (17,18) is as follows.
Starting from the initial homogeneous mixture of the two condensates, 
on the time scale given by 
\[
  t_I = 1/|\omega_{-max}| \; ,
\] 
domain patterns of the phase 
segregation with the characteristic length 
\[
  l_I = 1/q_{max}
\]
will appear.  Particularly, for the weakly segregated phase of 
$ \frac{G_{12}^2 }{G_{11}G_{22} } - 1 \rightarrow 0 $, we
have the length scale 
\[
  l_I^{-1} = \frac{\sqrt{m} }{\hbar} 
          \sqrt{ \frac{ 2\rho_{10}\rho_{20} G_{11} G_{22} }
                      { (\rho_{10} G_{11} + \rho_{20} G_{22} ) }
                 \left( \frac{G_{12}^2 }{G_{11}G_{22} } - 1 \right) }
        = \frac{ \sqrt{2} }{ \sqrt{ \Lambda_1^2 + \Lambda_2^2 } }  \; , 
\]
and for the strongly segregated phase  of
$ \frac{G_{12}^2 }{G_{11}G_{22} } - 1 \rightarrow \infty$,
we have 
\[
  l_I^{-1} = \frac{2 m^{1/2} }{\hbar} \left( \rho_{10}\rho_{20}G_{11}G_{22} 
               \left( \frac{G_{12}^2 }{G_{11}G_{22} } - 1 \right)\right)^{1/4}
   =\frac{\sqrt{2}}{ \sqrt{ \Lambda_1 \Lambda_2 } } \; . 
\]
Here  $ \Lambda_j = \xi_j /\sqrt{ {G_{12} }/{\sqrt{ G_{11}G_{22} } } - 1 }$
and $\xi_j = \sqrt{ {\hbar^2}/{2m_j} \; {1}/{ \rho_{j0}G_{jj} } }$ 
are the penetration and healing lengths in the binary BEC mixture 
\cite{ac}. 
Those length and times scales can be measured experimentally.
We will come back to the experimental situation below.

After the stage I of fast growth into the domain 
pattern characterized by the length scale $l_I$, 
the system will gradually approach to
the true ground state of the complete phase segregation: 
one condensate in one region and the second condensate in another region.
This stage is slow, dominated by the slowest mode, and is 
the subject of next section. 

It is now evident that the stage I of the growth of 
binary BEC's 
shares the same phenomenology of the initial stage of the classical
spinodal decomposition: the domination of the fastest growth mode, 
the appearance of domains of segregated phases, and
the conservation of particle numbers.
There are, however, two important differences.
First, the dynamical evolution of the binary BEC's is governed by
a coupled nonlinear time dependent Schr\"odinger equations,
not by a nonlinear diffusion equation supplemented with 
the continuity equation, the Cahn-Hilliard equation.
There is no external relaxation process  for the present wave functions.
Secondly, 
the energy of the binary BEC's is conserved during the growth process,
not as in the case of classical spinodal decomposition where
the system energy always decreases.
  
We wish to point out that the initial stage of 
spinodal decomposition without diffusion constant has been explored 
in heavy ion collisions for Fermi systems, 
based on the kinetic equation approach \cite{hpr}.
Qualitatively similar results have been obtained there.
The study in alkali BEC mixtures should
shed light on this important process, because the observation of real time 
evolution is in principle attainable experimentally.

\section{ Stage II: Merging and Oscillating between Domains }

The BEC binary mixture occurs in a trap.
This finite size effect of the droplet
leads to the broken symmetry of the condensate
profiles \cite{ca}, 
which tends to separate the condensates in mutually isolated regions.
This implies that 
there is no contact between different domains of the same
condensates formed in the stage I.
The classical spinodal process involves diffusion. An estimate of the
diffusion constant for the BEC system can be made from kinetics theory.
The ratio of the time scales for quantum and classical particle
transport is of the order of the ratio of the BEC cloud size
to the de Broglie wavelength.
This is much smaller than one for the experimental systems of interest.
The classical diffusion process is thus not important.
Because of the domains are not connected 
and because the diffusion for the BEC mixture is
extremely low, all the mechanisms for the late stage classical
spinodal decomposition are not applicable. 
We propose that it is the Josephson effect that is responsible for the
approach to equilibrium in stage II.
Two models for the Josephson effect, the `rigid pendulum' model 
\cite{tunneling1}  and 
the `soft pendulum' model \cite{tunneling2} will be discussed.
They both give the same time scale when the `Rabi' frequency is small.
Those model studies are performed for a single BEC condensate separated by
an external potential barrier. 
We adapt their analysis to the present situation
of one condensate sitting in the potential wells formed
by another condensate during the phase segregation process.

For a classical spinodal decomposition in a binary fluid mixture, 
there are several scenarios of growth in  stage II.
(1) For the domains to grow by the diffusion of materials
between them, a finite diffusion constant is needed.
For the binary BEC mixture, the temperature is too low for the
diffusion to occur. The diffusion constant is practically zero.
Furthermore, particles of one condensate 
have to diffuse into the domains of the other condensate
because domains of the same condensate are not connected.
This requires a finite activation energy.
Hence this process is not possible at low enough temperatures.
(2) The same argument also rules out the applicability of the
diffusion-enhanced collisions by which the diffusion field around the 
domains leads to an attraction between them.
(3) Another mechanism is the noise-induced growth. However, 
there is no external noise in the BEC described by the 
time dependent nonlinear Schr\"odinger equations. 
This mechanism is thus irrelevant here.
(4) A seemingly relevant one is the hydrodynamic growth 
that is driven by the pressure difference between the points of 
different curvatures. This would be a rather fast process. 
However, this also requires a connected phase of one of the species.
Hence, at zero temperature and for a finite system, classical
scenarios for the later stage will not work here.
We need to find a scenario working under the present conditions, and 
at the same time that is consistent with the time 
dependent nonlinear Schr\"odinger equations.

The only scenario which can transport particles across
a forbidden region at zero temperature is the tunneling process.
Let us consider the specific case of two domains of 
 condensate 1 separated by a domain with width $d$ of  condensate 2.
The ability of condensate 1 to tunnel through condensate 2
is described by the penetration depth $\Lambda$, 
as discussed in Ref. \onlinecite{ac}.
Hence the probability of condensate 1 to tunnel through condensate 2 
can be estimated as
\be
   p =   e^{ - 2 \frac{d}{\Lambda} }  \; ,
\ee
when $p$ is much smaller than 1.
The finiteness, though small, of the tunneling probability suggests 
that it is the Josephson effect responsible for the relaxation process in
the stage II. 
The Josephson effect 
leads to the merging of two domains of the same condensate at 
sufficiently low temperatures.
The dynamics of such motion may be governed by the `rigid pendulum' 
Hamiltonian for a Josephson junction \cite{tunneling1}:
\[
   H(\phi, n ) = E_J ( 1- \cos \phi ) + \frac{1}{2} E_C \;  n^2 \; ,
\]      
where $E_J$ is the Josephson coupling energy
determined by the tunneling probability,
$n= (n_x - n_y )/2$ is the particle number difference between 
the numbers of particles, $n_x$ and $n_y$, in the two domains,
and $E_C \equiv \partial \mu /\partial n $  is the `capacitive' energy
due to interactions. 
In the absence of external constraints, $ \mu = E_C n $.
The phase difference $\phi$ between the two domains is conjugated to $n$,
as in usual Josephson junctions.
Under the appropriate condition, such as low temperature and smallness of the
capacitive energy,  
there may be an oscillation  between the two domains of  condensate 1
separated by condensate 2.
In such a case, we may estimate that 
the oscillation period $ \tau_{II} = 2\pi/\omega_{JP} $, with the so-called
Josephson plasma frequency\cite{tunneling1}
\be
   \omega_{JP} =  \frac{ \sqrt{ E_C E_J }}{\hbar} \; .
\ee
For small tunneling probability, the Josephson junction energy may be
estimated as\cite{tunneling1}
\[
   E_J = n^{1/3}_{T} \hbar\omega_0 \;  e^{ - 2 \frac{d}{\Lambda} } \; ,
\]
and the capacitive energy as
\[
   E_C = \frac{2}{5} \left( \frac{n_T }{2} \right)^{-0.6} 
          \left(\frac{15a_{11} }{a_{0} }\right)^{0.4} {\hbar\omega_{0}} \; .
\]
Here $\omega_0$ is the harmonic oscillator frequency for  condensate 1
     in a harmonic trap, 
     $a_0 =  \sqrt{\hbar/m_1\omega_0} $ is the corresponding oscillator length,
     the $a_{11}$ is the scattering length of  condensate 1, and 
     $n_T = n_x + n_y$ is the total number of particles in domain $x$ and $y$.
Then the oscillatory time scale between the domains determines by 
the Josephson plasma frequency 
\be
   \tau_{II}^{-1} = \left( \frac{2a_0 }{15a_{11} \; n_T } \right)^{2/15} 
      \frac{\omega_0}{2\pi}  \;  e^{ - \frac{d}{\Lambda} }\; .
\ee
The rigid pendulum Hamiltonian would give a good description when
$n << n_T$.

Another description of the Josephson effect uses the 
`soft pendulum' Hamiltonian proposed in Ref. \onlinecite{tunneling2}:
\[
      H(\phi, n ) = - E_J(n) \cos \phi + \frac{1}{2} E_C \;  n^2 \; ,
\]
with the number dependent Josephson coupling energy
$ E_J(n) = n_T \hbar \omega_{R} \sqrt{1- (n/n_T)^2 } $.
Here $\omega_R\approx E_J/\hbar$ is the Rabi frequency determined by 
the overlap integral
between the two wave functions in the absence of the mutual repulsive 
interaction $G_{12}$.
The appropriate frequency scale in this case is
\be
   \omega^2_{sp} = \omega_{JP}^{2} + \omega_{R}^2 \; ,
\ee
which determines the time scale for the two domains to merge.
The Josephson plasma frequency in Eq.(22) of the soft pendulum model 
is  $\omega_{JP} = \sqrt{n_T E_c\omega_R/\hbar} \;  $.
For the experimental situation of interest, $\omega_R<<\omega_{JP}$.
Thus the two approaches give essentially the same answer. The detailed
derivation of $E_J$ was given in Ref. \onlinecite{tunneling1}.


Given the Josephson effect is the dominant
mechanism in the stage II, 
the time scale  to arrive at the ground state
will be determined by the Josephson effect at the final two domains, 
in which the two domains of condensate 1 is separated by 
the domain of  condensate 2, in the manner of $1 \, 2 \, 1 \, 2 $
spatial configuration 
for the case of equal numbers of the condensates.
The width of each domain is then $D/4$, with $D$ the size of the trap.  
According to the above analysis, 
the largest time scale determined by Eq.(22), the slowest mode, 
in the stage II is:
\be
   t_{II} = 2\pi/\omega_{sp}  \;  . 
\ee
The arrangement of $1\, 2 \, 1 $ spatial configuration 
may also occur here, in which it is
more likely for  condensate 2 to tunnel through 1 to the edge of the trap,
because of the larger tunneling probability.

\section{ Discussions }

The first question is that the quantum spinodal decomposition 
discussed above can happen or not.
In terms of the atomic scattering lengths of condensate atoms
$a_{jj}$, the interactions are 
$G_{jj} = {4\pi \hbar^2 a_{jj} }/{m_j } $.
The typical value of $a_{jj}$ for $^{87}$Rb is 
about $50$\AA.
The typical density realized for the binary BEC mixture is 
about $\rho_{i0} \sim 10^{14}/cm^3$.
Hence the healing length is 
$\xi = \sqrt{  ({\hbar^2}/{2m}) ({1}/{G_{jj} \rho_{i0} } ) }
     = \sqrt{1/(8\pi a_{jj}\rho_{i0} ) } \sim 3000$\AA.
For the different hyperfine states of $^{87}$Rb,
it is known now\cite{jila2,ca,ac} that 
${G_{12} }/{ \sqrt{ G_{11}G_{22} } } > 1$.
Hence the ground state of the phase segregated phase can be realizable.
Therefore, the quantum spinodal decomposition can happen.

Experimentally, starting from the initially homogeneous state,
after a short period of time 
a domain pattern does appear.
Then a damped oscillation between the domain pattern has been observed.
Eventually the binary BEC mixture sets to the segregated phase \cite{jila2}.
If we take ${G_{12} }/{ \sqrt{ G_{11}G_{22} } } = 1.04$,
 the penetration depth is 
$\Lambda = \xi/\sqrt{ {G_{12} }/{ \sqrt{ G_{11}G_{22} } } - 1 }  
     \sim 1.5 \mu$m.
The length scale $l_I$ determined by Eq.(17) in the stage I 
is $ 1.5 \mu$m, which is the same order of magnitude in comparison with
experimental data\cite{jila2}.
This length is also comparable to the domain wall width seen
experimentally.
The corresponding time scale $t_I$ (Eq. (18)) is then 6 ms, again the 
same order of magnitude, 
though both estimated $l_I$ and $t_I$ are somewhat smaller.
If ${G_{12} }/{ \sqrt{ G_{11}G_{22} } } = 1.0004$, the corresponding 
length and time scales in stage I according to the present analysis are
$ l_I = 15 \mu$m and $ t_I = 600$ms, respectively.
Time and length scales calculated in this way are 
larger than the numbers extracted from the experiment \cite{jila2}.
If we assume the damped oscillation to equilibrium observed experimentally
is the stage II discussed here, 
taking the Rabi frequency $\omega_R = 0.2 $nK = 4 Hz \cite{tunneling2} 
and the total number of particle $N_1 = 10^{6}$,
we find the period according to Eq. (23) is 30 ms, comparable to the 
experimental value. The estimation from Ref. \onlinecite{tunneling1}
gives a larger value of about 200 ms.
We think this estimate of the larger time scale is due to the use of a
larger estimate of $d/\Lambda$. 
In our view, d and $\Lambda$ are comparable in stage I. Thus for stage
II, $d/\Lambda$ is of the order of $n_d$, the number of domains formed
in stage I. For the experiment of interest to use, $n_d\approx 2$.
Thus for the case at hand, we think the first estimate is more
reasonable.

The present analysis shows that the stage I is insensitive to damping.
Similar conclusion has been reached in the study of heavy ion collisions
for fermions \cite{hpr}.
At this moment we do not have a reliable estimation of the damping
in the Josephson oscillation, as indicated experimentally.
Nevertheless, 
given the uncertainty in the value of $G_{12} $ or $a_{12}$,
we conclude that the stage II of the quantum spinodal decomposition
may have been observed.


\section{Conclusion}

In the present paper we have studied the problem of 
the dynamical evolution of the binary BEC mixtures starting from the 
homogeneously mixed unstable state.
We have found a parallel analogy to the usual spinodal decomposition
process by calling the present one the quantum spinodal decomposition. 
The quantum spinodal decomposition consists of two stages:
the stage I of initial fast growth of domains, characterized by 
a time and a length scale, and the stage II of 
a slow relaxation towards equilibrium dominated by the Josephson effect.
The coupled non-linear Schr\"odinger equations 
provide a good theoretical description, which 
enable us to calculate the time and length scales.
In comparison with recent experiments, we conclude that 
the quantum spinodal decomposition can be, and may have been,
realized.

The length and
time scales for Rb mixtures is controlled by the difference of
$r=G_{12}/\sqrt{G_{11}G_{22}}$ and 1. Since $r$ is close to 1
experimentally, the data can provide for a very sensitive estimate of
r-1. Since this parameter determines $\bf both$ the time and length
scale of stage I, it is a self-consistent check on the physical picture
provided here.

Periodic-like structures have also been observed in the phase
segregation of spin 1 Na mixtures \cite{stenger}. We think a similar
picture of quantum spinodal decomposition applies to that case as well.

{\ }

\noindent
{ This work was supported in part  by a grant from NASA (NAG8-1427). 
  One of us (PA) thanks H. Heiselberg and C.J. Pethick for 
  discussions, and the Bartol Research Institute as well 
  as the Department of Physics at University of Delaware for the pleasant 
  hospitality, where the main body of the work was completed. }
  We appreciate the D. S. Hall for
    sending us their data and informing us on measurement details.


\begin{thebibliography}{99}

\bibitem{cahn}
   J.W. Cahn, Tans. Met. Soc. AIME {\bf 242}, 166 (1968);

    J.S. Langer, in {\it Solids Far From Equilibrium}, ed. C. Godr\`eche
     (Cambridge University Press, Cambridge, 1992).
\bibitem{jila1}
 C.J. Myatt, E.A. Burt, R.W. Christ, E.A. Cornell, and C.E. Wieman, 
    Phys. Rev. Lett. {\bf 78}, 586 (1997). 
\bibitem{nlse}
   O. Penrose,  Phil. Mag. {\bf 42}, 1373 (1951);
   E.P. Gross, Nuovo Cimento {\bf 20}, 454 (1961);
   L.P. Pitaevskii, Sov. Phys. JETP {\bf 13}, 451 (1961);
   E. Demircan, P. Ao, and Q. Niu, Phys. Rev. {\bf B54}, 10027 (1996).
\bibitem{ac}
  P. Ao and S.T. Chui, Phys. Rev. {\bf A58}, 4836 (1998)(cond-mat/9809195).   
\bibitem{pu} 
 H. Pu and N.P. Bigelow, Phys. Rev. Lett. {\bf 80}, 1134 (1998); 
  and references therein.
\bibitem{ca}
  S.T. Chui and P. Ao, Phys. Rev. {\bf A59}, 1473 (1999).
\bibitem{hpr}
  H. Heiselberg, C.J. Pethick, and D.G. Ravenhall, 
   Ann. Phys. (NY), {\bf 223}, 37 (1993).
\bibitem{tunneling1}
  I. Zapata, F. Sols, and A.J. Leggett,
    Phys. Rev. {\bf A57}, R28 (1998).
\bibitem{tunneling2}
  A. Smerzi, S. Fantoni, S. Giovanazzi, and S.R. Shenoy, 
   Phys. Rev. Lett. {\bf 79}, 4950 (1997).
\bibitem{jila2}
 D.S. Hall, M.R.M. Matthews, J.R. Ensher, C.E. Wieman, and E.A. Cornell, 
    Phys. Rev. Lett. {\bf 81}, 1539 (1998). 
\bibitem{stenger}
   J. Stenger, S. Inouye, D.M. Stamper-Kurn, H.-J. Miesner, A.P. Chikkatur, 
   and W. Ketterle, Nature {\bf 396}, 345 (1998).  
\end{thebibliography}
\end{document}